%% file: main.tex
\documentclass[aip,preprint,amsmath,amssymb,floatfix,jcp]{revtex4-2}

\usepackage{graphicx}
\graphicspath{{figures/}}
\usepackage{physics}
\usepackage{amsmath}
\usepackage{multirow}
\usepackage{xcolor}
\usepackage{caption}
\usepackage{subcaption}
\usepackage{xcolor}
\usepackage{bm}
\usepackage{eufrak} 
\usepackage{amssymb} 
\usepackage[title,titletoc]{appendix}
\usepackage{diagbox}
\usepackage{dcolumn} 
\usepackage{siunitx} 

\usepackage{hyperref} 
\hypersetup{colorlinks=true, citecolor=blue, urlcolor=blue, linkcolor=blue}
\usepackage{cleveref}
  	\crefname{figure}{Figure}{Figures}
  	\crefname{table}{Table}{Tables}
  	\crefname{equation}{Eq.}{Eqs.}
  	\crefname{section}{Section}{Sections}
  	\crefname{subsection}{Section}{Sections}
  	\crefname{subsubsection}{Section}{Sections}
  	\crefname{algorithm}{Algorithm}{Algorithms}

\usepackage[section,frozencache]{minted}      

\usepackage{epstopdf}
\AppendGraphicsExtensions{.tiff}

\newcommand{\code}[1]{\texttt{#1}}

\usepackage{todonotes}

\newcommand{\pluseq}{\mathrel{+}=}

\begin{document}

\title{3-center and 4-center 2-particle Gaussian AO integrals on modern accelerated processors} 

\author{Andrey Asadchev}

\author{Edward F. Valeev}
\email{efv@vt.edu}
\affiliation{Department of Chemistry, Virginia Tech, Blacksburg, VA 24061}

\date{\today}

\begin{abstract}
  We report an implementation of the McMurchie-Davidson (MD) algorithm for 3-center and 4-center 2-particle integrals over Gaussian atomic orbitals (AOs) with low and high angular momenta $l$ and varying degrees of contraction for graphical processing units (GPUs). This work builds upon our recent implementation of a matrix form of the MD algorithm that is efficient for GPU evaluation of 4-center 2-particle integrals over Gaussian AOs of high angular momenta ($l\geq 4$) [{\em J. Phys. Chem. A} {\bf 127}, 10889 (2023)]. The use of unconventional data layouts and three variants of the MD algorithm allow to evaluate integrals in double precision with sustained performance between 25\% and 70\% of the theoretical hardware peak. Performance assessment includes integrals over AOs with $l\leq 6$ (higher $l$ is supported). Preliminary implementation of the Hartree-Fock exchange operator is presented and assessed for computations with up to quadruple-zeta basis and more than 20,000 AOs. The corresponding C++ code is a part of the experimental open-source \code{LibintX} library available at \textbf{\url{github.com:ValeevGroup/LibintX}}.
 \end{abstract}

\maketitle 

\section{Introduction}\label{sec:intro}
Evaluation of integrals of 1- and 2-particle operators over Gaussian atomic orbitals (AOs) is a performance-critical component of the vast majority of electronic structure computations on molecules and, increasingly, on periodic solids. This is due to the unparalleled efficiency of AO-based numerical representation for treating the low-energy physics of atomistic matter. Emergence of massively data-parallel processors such as general-purpose graphical processing units (GPGPUs or, simply, GPUs) as the primary computational platform for scientific computation made the development of efficient Gaussian AO integral kernels for such platforms necessary. Following the pioneering work by Yasuda 17 years ago,\cite{VRG:yasuda:2007:JCC} and the leading efforts by the \code{TeraChem} team,\cite{VRG:ufimtsev:2008:JCTC,VRG:ufimtsev:2009:JCTC} several GPU-capable AO integrals engines have been developed.\cite{VRG:asadchev:2010:JCTC,VRG:miao:2013:JCTC,VRG:yasuda:2014:IJQC,VRG:miao:2015:JCTC,VRG:rak:2015:CPL,VRG:kalinowski:2017:JCTC,VRG:song:2016:JCTC,VRG:kussmann:2017:JCTC,VRG:tornai:2019:JCTC,VRG:barca:2020:JCTC,VRG:barca:2020:2SICHPCNSAS,VRG:barca:2021:JCTC,VRG:barca:2021:PICHPCNSA,VRG:johnson:2022:JCTC,VRG:galvezvallejo:2022:MP,VRG:asadchev:2023:JCTC,VRG:asadchev:2023:JPCA,VRG:wu:2024:} Unfortunately, like \code{TeraChem}, most GPU integral engines are commercial or closed source, with \code{QUICK} library\cite{VRG:miao:2013:JCTC,VRG:miao:2015:JCTC} being a notable exception.

Although many GPU integral engines have been reported to date, there are strong similarities between most of them. The majority of the engines use recurrence-based integral evaluation strategies, albeit the Rys quadrature\cite{VRG:dupuis:1976:JCP} scheme is also employed;\cite{VRG:asadchev:2010:JCTC} most engines also use dedicated code generators to manage the implementation complexity of the complex recurrences. Several engines use mixed precision (FP64/FP32) to increase efficiency.
Lastly, and most importantly for this work, the performance of most GPU integral engines usually falls drastically for integrals over d and higher-$l$ AOs. For example, Barca et al. reported\cite{VRG:barca:2021:JCTC} a 2\% fraction of the hardware FP64 peak  throughput for evaluation of the [dd$\vert$dd] integrals using their engine implementing the Head-Gordon-Pople\cite{VRG:obara:1986:JCP,VRG:head-gordon:1988:JCP} refinement of the Obara-Saika\cite{VRG:obara:1986:JCP} scheme; contrast this to the 20\%-50\% fraction of the peak achieved for integrals over lower $l$. Unfortunately, only few papers report detailed breakdown of integral engine performance (most papers report aggregate timings for complex algorithms like Fock matrix formation), thereby making it difficult to understand variation of performance of individual kernels with the angular momenta. Nevertheless, even aggregate timings can sometimes be used to understand how the performance drops with the angular momenta. For example, Johnson et al.\cite{VRG:johnson:2022:JCTC} observed significant loss of efficiency of the Coulomb matrix GPU code based on the McMurchie-Davidson (MD) formalism\cite{VRG:mcmurchie:1978:JCP} compared to the CPU counterpart as the basis set was extended with high-$l$ functions. Specifically, the GPU vs single CPU core speedup in the cc-pV\{D,T,Q\}Z series reduced from $\sim50$ to $\sim10$ for the largest test cited in Figure 9 of Ref. \citenum{VRG:johnson:2022:JCTC}. The GPU-vs-CPU relative performance decrease is most likely due to the decrease of the absolute performance of the GPU engine with $l$ rather than due to the increased absolute performance of the CPU counterpart.

To address the drastic drop-off in performance with the angular momentum some have advocated\cite{VRG:kussmann:2017:JCTC} to use hybrid evaluation where the CPU engine is only used for integrals over high-$l$ AOs, with the GPU engine only used for low $l$. This is not generally viable due to the high-$l$ AO integrals accounting for a large fraction of the computational cost when high-$l$ basis sets are used, such as in medium- to high-precision computations on systems with light elements or in even low-precision computations on systems with heavy elements. Another possible solution is to fully deploy density fitting approximation, thereby avoiding the 4-center 2-electron integrals in lieu of 2- and 3-center 2-electron integrals. Unfortunately, straightforward use of density fitting can make it difficult or impossible to attain optimally-reduced complexity. Furthermore, robust density fitting basis sets are not available for wide swaths of the Periodic Table. Lastly, even when density fitting is an option the fitting basis sets can include functions with very high angular momenta ($l \gg 6$) thereby posing performance challenges. Thus it is currently not possible to completely eliminate the need for 4-center integral evaluation. Furthermore, evaluation of 3-center integrals over high-$l$ AOs needed for computations on heavy-element-containing compounds is likely to suffer from similar performance issues as the 4-center counterpart.

Recently we demonstrated\cite{VRG:asadchev:2023:JPCA} that it is possible to evaluate integrals over high-$l$ AOs ($l \geq 6$) efficiently on GPUs by deploying a matrix formulation of the McMurchie-Davidson scheme. Namely, our GPU-based implementation demonstrated  up to $\times15$ higher efficiency for the [ii$\vert$ii] integrals compared to the CPU-based evaluation using the Head-Gordon-Pople scheme (known to be significantly more FLOP optimal than the MD scheme). Note that the efficiency of the matrix form of the MD scheme for high-$l$ AO integrals was first demonstrated by Neese\cite{VRG:neese:2022:JCC} in the CPU context (low-$l$ integrals were still best computed using conventional Obara-Saika kernels), and are strongly related in spirit to the MD-based J matrix engine developments.\cite{VRG:shao:2000:CPL,VRG:williams-young:2023:JCP} The greatly increased vector unit width and smaller per-core fast memory on the modern GPUs compared to the CPU make the matrix form of the MD scheme even more beneficial and perhaps even optimal for the entire spectrum of AO integrals evaluated on GPUs.
Here we indeed show that the refinement of the matrix form of the MD scheme can be used to efficiently evaluate 3- and 4-center integrals over low- and high-$l$ AOs, with varying degrees of contraction, on GPUs even when compared to the more cost-effective Obara-Saika approach on CPU.

Computer implementation of the proposed evaluation approaches is publicly available as part of the \code{Libint} e\code{X}perimental (\code{LibintX}) open-source package.
The library aims to be a general-purpose engine for evaluation of operators over Gaussian AOs. Due to the large variation in the anatomy of existing Gaussian AO basis sets the library's focus is narrowed to (a) efficient support of segmented-contracted basis sets (hence, no special provisions are made for sp shells in the Pople basis set family), that (b) contain {\em only} solid harmonic AOs. These are very mild restrictions as they are met by the vast majority of practical computations.
The library also includes the density-fitting-accelerated Coulomb potential (J-engine) reported previously \cite{VRG:williams-young:2023:JCP}.
The library code is pure C++17 and CUDA. No external code generation is used; instead we rely on C++ metaprogramming techniques to ``generate'' compiler-friendly kernels, thereby keeping the source code compact and the compilation times low. This work is a major step towards having a portable integral engine that can execute on modern accelerators from NVIDIA and AMD, as well as CPUs with wide SIMD vectors.

The rest of the manuscript is structured as follows. \Cref{sec:formalism} recaps the McMurchie-Davidson (MD) scheme. \Cref{sec:implementation} describes several MD variants aimed at different classes of integrals. \Cref{sec:code} surveys the code organisation and API to help the readers interested in using  the \code{LibintX} source code.
\Cref{sec:performance} reports performance microbenchmarks for the new MD variants. \Cref{sec:exchange} describes a preliminary implementation of the exchange operator and initial performance benchmarks. \Cref{sec:summary} contains the summary of our findings and outlines future developments.

\section{Formalism}
\label{sec:formalism}

We start with a quick review of the standard McMurchie-Davidson (MD) scheme for the evaluation of 2-electron integrals over Gaussian AOs; the reader is referred to the original reference\cite{VRG:mcmurchie:1978:JCP} and other sources\cite{VRG:helgaker:2000:,VRG:samu:2018:} for additional details.

Our notation closely follows that of Obara and Saika.\cite{VRG:obara:1986:JCP}
An uncontracted primitive Cartesian Gaussian with exponent $\zeta_a \in\mathbb{R}^+$ and non-negative integer Cartesian ``quanta'' $\mathbf{a} \equiv \{a_x, a_y, a_z \}$ centered at $\mathbf{A} \equiv \{A_x, A_y, A_z\}$ is denoted by
\begin{align}
    \phi_\mathbf{a} (\mathbf{r}) \equiv x_A^{a_x} y_A^{a_y} z_A^{a_z} \exp(-\zeta_a r_A^2),
\end{align}
with $\mathbf{r}_A \equiv \{x_A, y_A, z_A\}, x_A \equiv x - A_x$, etc.
$l_\mathbf{a} \equiv a_x + a_y + a_z \geq 0$ is colloquially referred to as the ``angular momentum'' of a Gaussian.
3- and 4-center 2-electron Coulomb integrals over primitive AOs, denoted as
\begin{align}
\label{eq:def-acd}
[\mathbf{a}|\mathbf{c} \mathbf{d}] \equiv & \iint \frac{\phi_a(\mathbf{r}_1) \phi_c(\mathbf{r}_2) \phi_d(\mathbf{r}_2)}{\vert\mathbf{r}_1 - \mathbf{r}_2\vert} \, \mathrm{d}\mathbf{r}_1 \, \mathrm{d}\mathbf{r}_2 , \\
\label{eq:def-abcd}
[\mathbf{a} \mathbf{b}|\mathbf{c} \mathbf{d}] \equiv & \iint \frac{\phi_a(\mathbf{r}_1) \phi_b(\mathbf{r}_1) \phi_c(\mathbf{r}_2) \phi_d(\mathbf{r}_2)}{\vert\mathbf{r}_1 - \mathbf{r}_2\vert} \, \mathrm{d}\mathbf{r}_1 \, \mathrm{d}\mathbf{r}_2 .
\end{align}
will be also denoted generically as $[\mathrm{bra}\vert\mathrm{ket}]$.

The objective of this work is efficient evaluation
of 2-electron integrals over {\em contracted} Gaussians,
\begin{align}
    \phi_\mathbf{a} (\mathbf{r}) \equiv x_A^{a_x} y_A^{a_y} z_A^{a_z} \sum_{k}^{K_a} c_k \exp(-\zeta_{ak} r_A^2),
\end{align}
with $K_a \geq 1$ denoted by contraction {\em degree}.
We will not introduce separate notation for contracted integrals; we will highlight the appearance of contracted integrals only when it's not obvious from the context.
The contraction degrees of bra and ket, $K_\mathrm{bra}$ and $K_\mathrm{ket}$, respectively, are products of the contraction degrees of the constituent AOs. $K=K_\mathrm{bra} K_\mathrm{ket}$ will denote the total contraction degree of the integral.

Hermite Gaussians,
\begin{align}
    \Lambda_\mathbf{\tilde{a}} ({\bf r}) \equiv \left(\frac{\partial}{\partial x_A}\right)^{\tilde{a}_x} \left(\frac{\partial}{\partial y_A}\right)^{\tilde{a}_y} \left(\frac{\partial}{\partial z_A}\right)^{\tilde{a}_z} \exp(-\zeta_a r_A^2),
\end{align}
are used in the MD scheme to (exactly) expand primitive Cartesian Gaussians and their binary products:
\begin{align}
\label{eq:cart2herm-1}
    \phi_\mathbf{a}({\bf r}) = & \sum_{\tilde{a}_x=0}^{\tilde{p}_x \leq a_x} E_{a_x}^{\tilde{p}_x}  \sum_{\tilde{p}_y=0}^{\tilde{p}_y \leq a_y} E_{a_y}^{\tilde{p}_y} \sum_{\tilde{p}_z=0}^{\tilde{p}_z \leq a_z} E_{a_z}^{\tilde{p}_z} \Lambda_\mathbf{\tilde{p}}(\mathbf{r}) \equiv \sum_\mathbf{\tilde{p}} E_\mathbf{a}^\mathbf{\tilde{p}} \Lambda_\mathbf{\tilde{p}}(\mathbf{r}), \\
\label{eq:cart2herm-2}
    \phi_\mathbf{a}({\bf r})\phi_\mathbf{b}({\bf r}) = & \sum_{\tilde{p}_x=0}^{\tilde{p}_x \leq a_x+b_x} \left(E_x\right)_{a_x b_x}^{\tilde{p}_x}  \sum_{\tilde{p}_y=0}^{\tilde{p}_y \leq a_y + b_y} \left(E_y\right)_{a_y b_y}^{\tilde{p}_y} \sum_{\tilde{p}_z=0}^{\tilde{p}_z \leq a_z + b_z} \left(E_z\right)_{a_z b_z}^{\tilde{p}_z} \Lambda_\mathbf{\tilde{p}}(\mathbf{r}) \equiv \sum_\mathbf{\tilde{p}} E_{\mathbf{a} \mathbf{b}}^\mathbf{\tilde{p}} \Lambda_\mathbf{\tilde{p}}(\mathbf{r}) ,
\end{align}
with Hermite Gaussian exponent and origin for the 1- and 2-center bra given by
$\zeta_p \equiv \zeta_a$, $\mathbf{P} \equiv \mathbf{A}$ and
$\zeta_p \equiv \zeta_a + \zeta_b$, $\mathbf{P} \equiv \frac{\zeta_a \mathbf{A} + \zeta_b \mathbf{B}}{\zeta_a + \zeta_b}$, respectively.
Hermite Gaussian used to expand the 2-center ket will be annotated by $\tilde{\mathbf{q}}$.

Coefficients of Hermite Gaussians in expansions \cref{eq:cart2herm-1,eq:cart2herm-2} factorize along Cartesian axes: $E_\mathbf{a}^\mathbf{\tilde{p}} \equiv \prod_{i={x,y,z}} E_{a_i}^{\tilde{p}_i} $, and $E_\mathbf{a b}^\mathbf{\tilde{p}} \equiv \prod_{i={x,y,z}}\left(E_i\right)_{a_i b_i}^{\tilde{p}_i} $. They are evaluated straightforwardly by recursion\cite{VRG:mcmurchie:1978:JCP}:
\begin{align}
\label{eq:E1rr}
    \left(E_x\right)_{a_x+1}^{\tilde{p}_x} = & \frac{1}{2\zeta_a} \left(E_x\right)_{a_x}^{\tilde{p}_x - 1} + (\tilde{p}_x + 1) \left(E_x\right)_{a_x}^{\tilde{p}_x + 1} \\
\label{eq:E2rr1}
    \left(E_x\right)_{a_x+1 \, b_x}^{\tilde{p}_x} = & \frac{1}{2\zeta_p} \left(E_x\right)_{a_x \, b_x}^{\tilde{p}_x - 1} + (P_x - A_x) \left(E_x\right)_{a_x b_x}^{\tilde{p}_x} + (\tilde{p}_x + 1) \left(E_x\right)_{a_x b_x}^{\tilde{p}_x + 1} \\
\label{eq:E2rr2}
        \left(E_x\right)_{a_x \, b_x + 1}^{\tilde{p}_x} = & \frac{1}{2\zeta_p} \left(E_x\right)_{a_x \, b_x}^{\tilde{p}_x - 1} + (P_x - B_x) \left(E_x\right)_{a_x b_x}^{\tilde{p}_x} + (\tilde{p}_x + 1) \left(E_x\right)_{a_x b_x}^{\tilde{p}_x + 1},
\end{align}
(with the obvious $y$ and $z$ counterparts).
Recurrence \cref{eq:E1rr} is bootstrapped by definitions
\begin{eqnarray}
\label{eq:E1-00}
\left(E_i\right)_{0 }^{0} & = & 1, \\
\label{eq:E1-aa}
E_{a_i}^{\tilde{p}_i} & = & 0  \qquad (\tilde{p}_i \notin [0, a_i]  \vee\mod(\tilde{p}_i + a_i,2)=1).
\end{eqnarray}
Similarly, recurrences \cref{eq:E2rr1,eq:E2rr2} are bootstrapped by
\begin{align}
\label{eq:E2-000}
\left(E_i\right)_{0 0 }^{0} = & \exp\left(-\zeta_a \zeta_b \left(A_i-B_i\right)^2/\left(\zeta_a + \zeta_b\right)\right),\\
\label{eq:E2-abp}
E_{a_i b_i}^{\tilde{p}_i} = & \, 0 \qquad (\tilde{p}_i \notin [0, a_i+b_i]).
\end{align}
An important consequence of \cref{eq:E2rr2,eq:E2-abp} is that
\begin{align}
  \label{eq:E2-abab}
E_{a_i b_i}^{a_i+b_i} = & \frac{1}{\left(2\zeta_p\right)^{a_i+b_i}} E^0_{00}.
\end{align}
This important fact allows to reduce the cost of the transformation from Hermite Gaussian to Cartesian Gaussian basis, by recognizing that all Cartesian integrals with the same $a_i+b_i$ share the same contribution from the Hermite Gaussian with quantum number $\tilde{p}_i = a_i+b_i$. Evaluation of this contribution can occur on the fly due to the extremely simple form of \cref{eq:E2-abab}.
This effectively allows to implement the Hermite-to-AO transformation with the total angular momentum of bra/ket reduced by 1, which can produce significant savings for low and medium angular momenta (e.g., resulting in savings of $\sim43\%$ for the $\vert$dd] case). For high angular momenta the savings due to \cref{eq:E2-abab} decrease and have less effect overall.
Similarly to the on-the-fly evaluation of contributions from \cref{eq:E2-abab} to the 2-center Hermite-to-AO transformation, the absence of geometric prefactors in the 1-center $E$ coefficient recurrence (\cref{eq:E1rr}) allows to fuse their evaluation into the Hermite-to-AO transformation for {\it all} values of Hermite Gaussian quanta.

Since most of the time we are interested in evaluation of integrals over (real) {\em solid harmonic} Gaussian AOs, and for $l\geq 2$ solid harmonics are less numerous than Cartesians, matrices $E$ are first contracted with the Cartesian-to-solids coefficient matrices\cite{VRG:schlegel:1995:IJQC} to produce matrices $H$ that transform from (primitive) Hermite Gaussians to (primitive) real solid Gaussians directly:
\begin{align}
\label{eq:solid2herm-1}
    H_\mathbf{a}^{\mathbf{\tilde{p}}} = & \sum_{\mathbf{p}} C_{l_a m_a}^{\mathbf{a}} E^{\mathbf{\tilde{p}}}_{\mathbf{a}}, \\
\label{eq:solid2herm-2}
    H_\mathbf{a b}^{\mathbf{\tilde{p}}} = & \sum_{\mathbf{a}} C_{l_a m_a}^{\mathbf{a}} \sum_{\mathbf{b}} C_{l_b m_b}^{\mathbf{b}} E^{\mathbf{\tilde{p}}}_{\mathbf{a b}},
\end{align}
where $C_{l_a m_a}^{\mathbf{a}}$ is the coefficient of Cartesian Gaussian $\phi_\mathbf{a}$ in solid harmonic Gaussian $\phi_a$ with angular momentum quanta $l_a$ and $m_a$; for the detailed definitions see  Ref. \citenum{VRG:schlegel:1995:IJQC}.

{\em Shells} are groups of solid harmonic and Hermite Gaussians that share exponents, origin, contraction coefficients, and orbital quantum numbers. There are $N(l) \equiv 2 l+1$ and $\tilde{N}(l) \equiv (l+1)(l+2)(l+3)/6$ solid harmonic and Hermite Gaussians in a shell of angular momentum $l$, respectively. Shells of angular momenta \{0, 1, 2, 3, 4, 5, 6\} will be denoted by \{s, p, d, f, g, h, i\}, or by their angular momentum itself. E.g., [12$\vert$34] and [pd$\vert$fg] will denote a {\em class} of integrals representing any {\em shellset} of 4-center integrals with p, d, f, and g shells on each respective center. The number of solid harmonic Gaussians in bra and ket shellsets will be denoted by $N_\mathrm{bra} = N(l_\mathrm{bra})$ and $N_\mathrm{ket} \equiv N(l_\mathrm{ket}) $, respectively, with ${l_\mathrm{bra},l_\mathrm{ket}}$ the sum of angular momenta of bra and ket AOs. Similarly, the number of Hermite Gaussians needed to expand the bra and ket shellsets will be denoted by $\tilde{N}_\mathrm{bra} \equiv \tilde{N}(l_\mathrm{bra})$ and $\tilde{N}_\mathrm{ket} \equiv \tilde{N}(l_\mathrm{ket})$, respectively.

The use of \cref{eq:cart2herm-1,eq:cart2herm-2,eq:solid2herm-1,eq:solid2herm-2} allows to express AO integrals over contracted solid harmonic Gaussians as linear combinations
of the integrals between two primitive Hermite Gaussians:
\begin{align}
[\mathbf{\tilde{p}} | \mathbf{\tilde{q}}] \equiv & \,
\iint
\frac{\Lambda_\mathbf{\tilde{p}}(\mathbf{r}_1) \Lambda_\mathbf{\tilde{q}}(\mathbf{r}_2)}{\vert\mathbf{r}_1 - \mathbf{r}_2\vert} \, \mathrm{d}\mathbf{r}_1 \,\mathrm{d}\mathbf{r}_2.
\end{align}
These can be evaluated directly,
\begin{align}
\label{eq:pq}
[\mathbf{\tilde{p}} | \mathbf{\tilde{q}}] \equiv (-1)^{l_\mathbf{\tilde{q}}} [\mathbf{\tilde{p}}+\mathbf{\tilde{q}}]^{(0)},
\end{align}
from the auxiliary integral,
\begin{align}
[\mathbf{\tilde{r}}]^{(m)} \equiv \left(\frac{\partial}{\partial x_R}\right)^{\tilde{r}_x} \left(\frac{\partial}{\partial y_R}\right)^{\tilde{r}_y} \left(\frac{\partial}{\partial z_R}\right)^{\tilde{r}_z} [\mathbf{0}]^{(m)}.
\end{align}
$[\mathbf{0}]^{(m)}$ is related to the Boys function $F_m(x)$ (or similar quantities for non-Coulomb integrals\cite{VRG:ahlrichs:2006:PCCPP}):
\begin{align}
\label{eq:0m}
[\mathbf{0}]^{(m)} \equiv \, & (-2 \rho)^m \frac{2 \pi^{5/2}}{\zeta_p \zeta_q \sqrt{\zeta_p+\zeta_q}} F_m(\rho |\mathbf{P}-\mathbf{Q}|^2) , \\
F_m(x) \equiv \, & \int_0^1 \, \mathrm{d}y \, y^{2m} \exp(-x y^2), \label{eq:boys} \\
\rho \equiv \, & \frac{\zeta_p \zeta_q}{ \zeta_p + \zeta_q }.
\end{align}
The auxiliary integrals are evaluated recursively,
\begin{align}
\label{eq:r}
[\mathbf{\tilde{r}}+\mathbf{1}_i]^{(m)} = & \tilde{r}_i [\mathbf{\tilde{r}}-\mathbf{1}_i]^{(m+1)} + \left(P_i - Q_i\right) [\mathbf{\tilde{r}}]^{(m+1)},
\end{align}
starting from $[\mathbf{0}]^{(m)}$.
Note that the number of the 2-index Hermite integrals $[\mathbf{\tilde{p}} | \mathbf{\tilde{q}}]$ needed to compute a single shellset of uncontracted $[\mathrm{bra}|\mathrm{ket}]$ integrals, $\tilde{N}_b\tilde{N}_k$, typically greatly exceeds the number of the 1-index Hermite integrals $[\mathbf{\tilde{r}}]^{(0)}$, namely $\tilde{N} \equiv \tilde{N}(l_\mathrm{bra}+l_\mathrm{ket})$,  involved in \cref{eq:pq}.

The complete sequence of transformations involved in the MD scheme for evaluation of {\em primitive} 2-electron integrals $[\mathrm{bra}\vert\mathrm{ket}]$ is described in \cref{tab:md-scheme}. For each step we list the size of its input $A_\mathrm{in}$ and output $A_\mathrm{out}$ integrals, in words, and the number of floating-point operations (FLOPs), $N_\mathrm{F}$, involved in each step. Transformation from Hermite to AO basis assumes that the ket has higher $K$ and lower $l$ than the bra, hence the former is transformed first:
\begin{align}
\label{eq:ket-from-hermite}
[\tilde{\mathbf{p}}\vert\mathrm{ket}] = & \sum_{\tilde{\mathbf{q}}} [\tilde{\mathbf{p}}\vert \tilde{\mathbf{q}}] H_\mathrm{ket}^{\tilde{\mathbf{q}}}, \\
\label{eq:bra-from-hermite}
[\mathrm{bra}\vert\mathrm{ket}] = & \sum_{\tilde{\mathbf{p}}} [\tilde{\mathbf{p}}\vert \mathrm{ket}] H_\mathrm{bra}^{\tilde{\mathbf{p}}}.
\end{align}

In Ref. \citenum{VRG:asadchev:2023:JPCA} we demonstrated implementation of the scheme in \cref{tab:md-scheme} using general matrix-matrix (GEMM) kernels in the standard BLAS library. Whereas for integrals over $l\geq4$ AOs we saw excellent performance approaching 50\% of the hardware peak, for lower $l$ the lower efficiency indicated the need for alternative approaches. To understand the performance bottlenecks and the proposed implementation improvements we will model performance using Roofline,\cite{VRG:williams:2009:CA} a simple performance model that assumes every kernel reads $A_\mathrm{in}$ words, performs $N_\mathrm{F}$ FLOPs of work, and writes $A_\mathrm{out}$ words. For example, as illustrated by the $[66|66]$ integral, in the high-$l$ limit GEMM transformations $[\tilde{\mathbf{p}}\vert\tilde{\mathbf{q}}] \to [\tilde{\mathbf{p}}\vert\mathrm{ket}] \to [\mathrm{bra}\vert\mathrm{ket}]$ account for $>99\%$ of FLOPs. This explains why the matrix formulation {\em can} perform well for high-$l$ integrals. To determine whether a kernel will be memory bound (i.e., limited by the rate of reading/writing data) by providing an estimate of the arithmetic intensity $I$ for a given operation. If $I$ is greater than the ratio between the theoretical peak FLOPs per second and maximum bandwidth indicates balance indicates a compute-bound problem, and vice versa.  With data residing in the main memory the critical ratio for a PCIe V100 GPU is $\frac{7~\text{TFLOP/s}}{900~\text{GB/s}}$ = 7.7 FLOP/byte or 62 FLOPs per FP64 {\it word}. The arithmetic intensity for $[66|66]$ is well above 7.7 ratio, which correlates with its good performance. Already for medium-$l$ integrals (exemplified by $[33|33]$) the estimated intensity drops below the 7.7 ratio. For low-$l$ integrals the estimated intensities are far too low to expect good performance if the data resides in main memory. However, the Roofline threshold for computation on data in registers and/or scratchpad memory is substantially less than 1; good performance is possible even for low-$l$ integrals as long as the data stays resident in the highest levels of GPU memory hierarchy.
This can be done by fusing steps together such that:
\begin{itemize}
    \item for low-$l$ integrals all but the final results reside in registers;
    \item for medium-$l$ integrals 2-center Hermite integrals, $[\tilde{\mathbf{p}}\vert\tilde{\mathbf{q}}]$ are never read or written to main memory.
\end{itemize}
Thereby we arrive at 3 different variants of the MD scheme that we label by the number of Hermite Gaussians in the intermediate produced by the first step:
\begin{itemize}
    \item {\bf V2}: the original MD formulation of Ref. \citenum{VRG:asadchev:2023:JPCA}, with the $[\tilde{\mathbf{p}}\vert\tilde{\mathbf{q}}]$ and subsequent intermediates resident in the main GPU memory;
    \item {\bf V0}: formulation targeted at low-$l$ integrals, with only target $[\mathrm{bra}\vert\mathrm{ket}]$ integrals written to the main GPU memory;
    \item {\bf V1}: formulation targeted at medium-$l$ integrals, which fuses steps $[\tilde{\mathbf{r}}]^{(0)} \to [\tilde{\mathbf{p}}\vert\tilde{\mathbf{q}}]$  and $[\tilde{\mathbf{p}}\vert\tilde{\mathbf{q}}] \to [\tilde{\mathbf{p}}\vert\mathrm{ket}]$.
\end{itemize}
Details of these variants will be discussed next.

\begin{table}[]
    \centering
    \begin{tabular}{c|rrrrrrrrr}
    \hline
& \multicolumn{9}{c}{MD transformations} \\ \hline
& $[\mathbf{0}]^{(m)}$ &
$\overset{\mathrm{\cref{eq:r}}}{\to}$ & $[\tilde{\mathbf{r}}]^{(0)}$ &
$\overset{\mathrm{\cref{eq:pq}}}{\to}$ &$[\tilde{\mathbf{p}}\vert\tilde{\mathbf{q}}]$ &
$\overset{\mathrm{\cref{eq:ket-from-hermite}}}{\to}$ &$[\tilde{\mathbf{p}}\vert\mathrm{ket}]$ &
$\overset{\mathrm{\cref{eq:bra-from-hermite}}}{\to}$ &$[\mathrm{bra}\vert\mathrm{ket}]$ \\
\\
& \multicolumn{9}{c}{$[\mathrm{bra} \vert \mathrm{ket}]$} \\ \hline
$A_\mathrm{in}$ & & $L+1$ & & $\tilde{N}$ & & $\tilde{N}_\mathrm{bra} \tilde{N}_\mathrm{ket}$ & & $\tilde{N}_\mathrm{bra} N_\mathrm{ket} $ & \\
$A_\mathrm{out}$ & & $\tilde{N}$ & & $\tilde{N}_\mathrm{bra} \tilde{N}_\mathrm{ket}$ & & $\tilde{N}_\mathrm{bra} N_\mathrm{ket}$ & & $N_\mathrm{bra} N_\mathrm{ket}$ & \\
$N_\mathrm{F}$ & & $R_L$ & & $\tilde{N}_\mathrm{bra} \tilde{N}_\mathrm{ket}/2$ & & $2 \tilde{N}_\mathrm{bra} \tilde{N}_\mathrm{ket} N_\mathrm{ket}$ & & $2 N_\mathrm{bra}  \tilde{N}_\mathrm{bra} N_\mathrm{ket}$ \\
\\
& \multicolumn{9}{c}{$[11 \vert 11]$} \\ \hline
$A_\mathrm{in}$& & 5 & & 35 & & 100 & & 90 & \\
$A_\mathrm{out}$& & 35 & & 100 & & 90 & & 81 & \\
$N_\mathrm{F}$& & 396 & & 50 & & 1800 & & 1620 & \\
$I_\mathrm{FP64}$& & 1.2 & & 0.0 & & 1.2 & & 1.2 & \\
\\
& \multicolumn{9}{c}{$[33 \vert 33]$} \\ \hline
$A_\mathrm{in}$& & 13 & & 455 & & 7,056 & & 4116 & \\
$A_\mathrm{out}$& & 455 & & 7,056 & & 4,116 & & 2,401 & \\
$N_\mathrm{F}$& & 7,182 & & 3,528 & & 691,488 & & 403,368 & \\
$I_\mathrm{FP64}$& & 1.9 & & 0.1 & & 7.7 & & 7.7 & \\
\\
& \multicolumn{9}{c}{$[66 \vert 66]$} \\ \hline
$A_\mathrm{in}$& & 25 & & 2,925 & & 207,025 & & 76,895 & \\
$A_\mathrm{out}$& & 2,925 & & 207,025 & & 76,895 & & 28,561 & \\
$N_\mathrm{F}$& & 71,331 & & 103,514 & & 69,974,450 & & 25,990,510 & \\
$I_\mathrm{FP64}$& & 3.0 & & 0.1 & & 30.8 & & 30.8 & \\\hline
    \end{tabular}

$^\dagger$ $L = l_\mathrm{bra}+l_\mathrm{ket}$, $R_L = (2 + L) (7 + L) (12 + 5 L + L^2)/8$.

    \caption{Schematics of the MD evaluation of contracted 2-electron integrals and its rough performance model. For each evaluation step the corresponding memory footprints (in words) of input ($A_\mathrm{in}$) and output ($A_\mathrm{out}$) data, number of FLOPs ($N_\mathrm{F}$) listed for generic and concrete prototypical shellsets of integrals. For concrete shellsets the FP64 arithmetic intensity, $I_\mathrm{FP64}=N_\mathrm{F}/8(A_\mathrm{in}+A_\mathrm{out})$, in FLOP/byte, is also listed.$^\dagger$}
    \label{tab:md-scheme}
\end{table}

\section{McMurchie-Davidson Algorithm Variants}
\label{sec:implementation}

Narrowing the scope of our MD implementation to 3 variants, and targeting solid harmonic segmented Gaussian basis sets still leaves a number of possible design parameters  on the table.
To narrow down the design scope further we made several additional assumptions about the desired use cases as well as the target GPU platform parameters.
\begin{itemize}
\item The cost of MD depends on whether bra or ket is transformed  to the AO basis first; transforming the side with higher degree of contraction first is preferred. We assume input can be sorted such that ket has same or higher contraction order as bra.
\item The target GPU's main memory transaction size is 128 bytes (16 FP64 words), hence threads need to address the GPU main memory in groups of 16.
\item Two or more thread blocks must be resident on each streaming multiprocessor (SM) to hide memory and instruction latencies.
\item Target GPU has a register file of at least 256KB. This is true for all latest NVIDIA devices (P100, V100, A100, H100) and for the AMD MI100 device.
That corresponds to two or more thread blocks of 128 threads each, with up to 1024 bytes of registers per thread.
\item Target GPU has up to either 32KB (MI100,P100) or 48KB (static shared memory restriction on CUDA devices), or fast shared memory (LDS in AMD documentation) per thread block.
\item The engine capabilities must not be limited by the size of the basis set; no $N_{shells}^2$ data should be stored.
\end{itemize}

Each of the three MD variants is implemented as a sequence of one or more kernels, each performing one or more of the four MD transformations described in \cref{tab:md-scheme}.
To fully utilize the massively-parallel devices like modern GPUs many integral shellsets must be evaluated per launch. Unlike the popular CPU-focused Gaussian integrals libraries that compute one shellset at a time, \code{LibintX} must evaluate large batches of shellsets at a time. Hence it is important to discuss the details of the memory order of the intermediates and the final integrals. We will use $ij$ and $kl$ to refer to bra and ket shell pair indices, respectively.  Due to screening these index sets are {\it not} Cartesian products of two index sets in general. Shell pairs with the same angular momentum and same total contraction degrees can be grouped together, the values of exponents and coefficients do not matter. From here on $ab$ and $cd$ will refer to Cartesian products of Gaussian AOs in bra and ket, respectively. For the 3-center integrals $i$ and $a$ will denote bra shell and AO indices, respectively. $K_{b}$ and $K_{k}$ will denote primitive indices of bra and ket, respectively. $p$, $q$, and $r$ will index Hermite Gaussians of the $[\tilde{\bf p}|\tilde{\bf q}]$ and $[\tilde{\bf r}]^{(0)}$ intermediates.

The memory order of final integrals produced is $V[ij,ab,cd,kl]$, i.e., a 4-dimensional column-major array. Note that this order differs from that mentioned in our prior work \cite{VRG:asadchev:2023:JPCA}.
Integral batches can be accessed in conventional order via a slicing operation, eg $ij,kl$-th batch is accessed as $V[ij,:,:,kl]$ using NumPy syntax.

While there is sufficient and consistent source of fine-grained (thread-level) parallelism within a single high-$l$ shellset, it is not true for shellsets with lower $l$, the targets of variants {\bf V0} and {\bf V1}. In these cases, parallelism over shell pair indices, $ij$ or $kl$, is the only consistent source of fine-grain parallelism. Since the Hermite-to-AO transformation is done for the ket first, index $kl$ is chosen to be outermost to amortize the memory costs of the $H_\mathrm{ket}^{\tilde{\mathbf{q}}}$ matrices (\cref{eq:ket-from-hermite}) over multiple $ij$ indices (i.e., $H_\mathrm{ket}^{\tilde{\mathbf{q}}}$ corresponding to each ket shell pair $kl$ is used to evaluate $[\mathrm{bra}\vert\mathrm{ket}]$ for multiple bra shell pairs $ij$). This leaves the $ij$ index to be innermost, hence the peculiar memory order. Furthermore, to address the 128 byte memory transaction requirement the $ij$ indices need be processed in groups of $16n$ ($n=1,2,4,8$). The current GPU convention uses 3-dimensional $x,y,z$ thread blocks and index $ij$ is mapped onto the $x$ coordinate.

Bra and ket transformation matrices $H$ and Gaussian pair data $G$ are computed as needed for each set $ij$ and $kl$, never for the entire basis. To permit various optimisations, transpose, and slicing operations arrays $G$ and $H$ are stored contiguously in memory for each shell pair index and primitive index combination; i.e. $\{G,H[ab,p]\}[ij,K_{b}]$.

\subsection{{\bf V0} Variant}

The limiting factor in {\bf V0} is the limited register memory to store the half-transformed integrals, $[\tilde{\bf p}|\mathrm{ket}]$, with ket in contracted AO basis. For the aforementioned microarchitecture assumptions, a 256KB register file, split between 2 thread blocks and partitioned among the minimum of 16 $ij$ shell pair indices, affords $\sim1024$ FP64 words for their storage.  This number is further reduced since various temporaries, such as loop counters, need to reside in registers as well. This limits the {\bf V0} variant to [dd$\vert$dp] shellsets or smaller; [dd$\vert$dd] on other hand would result in excessive register spills.

Although some shellsets with high-$l$ bra {\em and} low-$l$ ket, such as [gg$\vert$ss], can be evaluated via {\bf V0} without register spills, the compute cost of ket is too low to amortize the cost of reading the $H^{\tilde{\bf p}}_\mathrm{bra}$ matrices. Such integral types are better suited for {\bf V1} or {\bf V2} variants.

A 128-thread block is assigned either 16, 32, or 128 $ij$ indices. The latter corresponds to a simple 1 thread to 1 $ij$ mapping. In the first two cases each thread is assigned single $ij$ and a subset of $p$ indices. The parallelization over only $ij,p$ indices guarantees that the innermost ket loops remain known at compile time and can be fully optimized by the compiler. 3-center integrals with $l_\mathrm{bra} \ll l_\mathrm{ket}$, such as [s$\vert$gg], are a corner case where parallelisation over $q$ is prefered.

\subsection{{\bf V1} Variant}
The {\bf V1} variant consists of three kernels, evaluating $[\tilde{\mathbf{r}}]^{(0)}$, $[\tilde{\mathbf{p}}\vert\mathrm{ket}]$ (with contracted ket), and fully-contracted $[\mathrm{bra}\vert\mathrm{ket}]$ integrals, laid out in memory as $[ij, r, kl]$, $[ij, p, cd, kl]$ and $[ij, ab, cd, kl]$, respectively. Note that the primitive indices are omitted for simplicity. The applicability of ${\bf V1}$ is limited by whether the $H^{\tilde{\bf q}}_\mathrm{ket}$ matrix can fit in the shared memory. Assuming the maximum of 48KB of shared memory per thread block, this allows to use ${\bf V1}$ for integrals with ket $\vert$gf]; to handle ket $\vert$gg] $120\times 9 \times 9 \times 8 = 78$KB of shared memory would be required.

{\bf V1} has the lowest arithmetic intensity of the three MD variants.  To improve its performance we employ various small optimizations, such as array tiling to maximize the cache locality and batching several primitives to amortize the cost of writing out contracted integrals (refer to \code{feature/gpu/md} branch for full source code). For small kets (e.g., [gg$\vert$ss]) the first two transformations are fused together, i.e., integrals $[\tilde{\mathbf{p}}\vert\mathrm{ket}]$ are computed directly in a single kernel (implemented as a variation of {\bf V0} kernel).

Lastly, {\bf V1} is not used for the evaluation of 3-center integrals.

\subsection{{\bf V2} Variant}
\label{sec:v2}

To produce the target integrals in the desired $[ij,ab,cd,kl]$ order, {\bf V2} differs from that reported previously.\cite{VRG:asadchev:2023:JPCA} The first kernel evaluates $[\tilde{\bf p}\vert\tilde{\bf q}]$ integrals in $[p,ij,q,kl]$ order, followed by transformations to AOs ($V$ refers to various integral arrays and their memory order):
\begin{eqnarray}
\label{eq:V2-eq1}
V[p,ij,cd,kl] & \pluseq & V[p,ij,q,kl] H[cd,q,kl] , \\
\label{eq:V2-eq2}
V[cd,kl,p,ij] & = & V[p,ij,cd,kl], \\
\label{eq:V2-eq3}
V[ab,cd,kl,ij] & \pluseq & H[ab,p,ij] V[cd,kl,p,ij] , \\
\label{eq:V2-eq4}
V[ij,ab,cd,kl] & = & V[ab,cd,kl,ij].
\end{eqnarray}
Steps \eqref{eq:V2-eq1} and \eqref{eq:V2-eq3} are batched GEMMs with the second argument matrix transposed; steps \eqref{eq:V2-eq2} and \eqref{eq:V2-eq4} are simple matrix transposes.

For the 3-center integrals evaluation of 2-index Hermite integrals and transformation of the bra to the AO basis is fused into a single kernel, producing $[\mathrm{bra}\vert\tilde{\bf q}]$ laid out as $[q,i,a,kl]$. Then ket is transformed to the AO basis using a single batched GEMM:
\begin{eqnarray}
V[i,a,cd,kl] & \pluseq & V[q,i,a,kl] H[cd,q,kl]
\end{eqnarray}

\section{Implementation in \code{LibintX} Library}
\label{sec:code}

The implementation of the above MD variants for 3- and 4-center 2-particle integrals is freely available at \textbf{\url{github.com:ValeevGroup/LibintX}} (branch \code{feature/gpu/md}) under terms of the LGPL3 license.
As in our previous\cite{VRG:asadchev:2023:JCTC,VRG:asadchev:2023:JPCA,VRG:williams-young:2023:JCP} work the library is wrtten in standard C++17 and its CUDA extension; there is no custom code generator. The entire 3- and 4-center GPU source code is fewer than 3000 lines. Library compilation for $l=6$ using 16 threads takes 13 minutes for 3-center integrals and 45 minutes for 4-center integrals.  The compilation time is a one-time cost:  the internal template instantiations as well as CUDA headers are {\em not} exposed to library interface.

The code is organized as follows:
\begin{itemize}
  \item \code{md.kernel.h}: GPU kernel code common to both (3- and 4-center) integral types;
  \item \code{md\{3,4\}.kernel.h}: GPU kernel code specific to 3- and 4-center integrals;
  \item \code{md\{3,4\}.kernel.cu}: instantiations of 3- and 4-center GPU kernels;
  \item \code{md\{3,4\}.\{h,cc\}}: 3- and 4-center host-side code.
\end{itemize}
The CMake build harness creates on the fly the object libraries corresponding to bra and ket angular momenta from the single \code{md\{3,4\}.kernel.cu} source file. For every valid combination $\{l_\mathrm{bra},l_\mathrm{ket}\}$ the three implementation variants are checked for {\em feasibility} in order {\bf V0}, {\bf V1}, {\bf V2}; the first variant that is deemed feasible is actually used for that integral class. Feasibility is determined by the shared memory and register requirements using a simple heuristic that does not require to generate GPU kernel first. Maximum shared memory size is a compile time parameter that can be used to tune for various GPUs.

\Cref{listing:md4} illustrates \code{LibintX}'s simple C++ API for computing 4-center integrals; the 3-center integral API is nearly identical. An analogous Python API, compatible with the standard \code{cuPy} module, is also provided.

\begin{center}
\begin{listing}[H]
\inputminted{c++}{c++api.h}
\caption{C++ API for computing 4-center integrals.}
\label{listing:md4}
\end{listing}
\end{center}

\section{Performance}
\label{sec:performance}

The performance of the new implementation on a GPU was assessed against the reference \code{Libint} library\cite{VRG:valeev:2021:libint-2.7.0} compiled with \code{-march=native -mtune=native -Ofast} flags.
Following the recent trend\cite{VRG:barca:2021:JCTC,VRG:asadchev:2023:JCTC,VRG:neese:2022:JCC,VRG:asadchev:2023:JPCA} we assess performance by {\em microbenchmarking} the integral kernels, i.e., we analyze their performance for specific integral classes to provide a more detailed picture of the performance.

The microbenchmarking setup was identical to that used in our previous work.\cite{VRG:asadchev:2023:JPCA}
Since the \code{Libint}
kernels evaluate single shellset at a time, each kernels was executed number of times corresponding to GPU problem size with the same arguments and buffer to eliminate
the cost of initialization.
The number of integrals for each \code{LibintX} invocation was chosen so that the final results fit under 500 MB of memory.
The performance does depend on problem dimensions to an extent; for example, small input or very lopsided $ij$,$kl$ inputs will result in poor performance.
The GPU timings include all steps, including evaluation of the $H$ and pair data.
Gaussian shells and centers were randomly generated to stress the interpolation-based Boys function code while avoiding screening effects.

\Cref{tab:speedup4} presents the observed speedup of the MD-based  evaluation of 4-center integrals on GPU using \code{LibintX} vs the Obara-Saika-based\cite{VRG:obara:1986:JCP,VRG:obara:1988:JCP} Head-Gordon-Pople\cite{VRG:head-gordon:1988:JCP} (OSHGP) counterpart on CPU using \code{Libint}.
The baseline for the comparison is the ratio of peak FLOP rate of the V100 GPU to that of the 1 Xeon Gold 6136 Skylake CPU {\it core} is
73:1; speedup greater than 73 indicates higher effective efficiency of the MD implementation.
Significant effort was spent to ensure high performance even for unfavorable cases such as [33$\vert$00].  Across integral classes and contraction orders the speedup remains higher than the 73:1 GPU:CPU performance ratio; there are dips in relative performance where, for example, OS scales sublinearly with respect to contraction order, which is noticeable for [gg$\vert$gg] and above.
At this time we chose not to pursue early contraction MD schemes.
For 3-center integrals (\cref{tab:speedup3}) the speedups are more modest, likely due to higher performance of the OSHGP implementation on the GPU. Nevertheless, other than few exceptions, the target speedup was reached in all cases.

The discrepancy between flop ratio and much higher performance ratio suggests that CPU MD is potentially faster overall than OSHGP.  In the near future we will present such CPU implementation that outperforms OSHGP and where speedups closely track the flop ratio.  There are cases where OSHGP does perform better, eg high-contraction high-angular momentum integrals, but such cases are rare with the basis sets we are targeting.

\begin{table}
  \centering
  \begin{tabular}{c|ccc}
    \hline
    Integral Class & \multicolumn{3}{c}{$\{K_\mathrm{bra},K_\mathrm{ket}\}$}  \\
    & $\{1,1\}$ & $\{1,5\}$ & $\{5,5\}$ \\
    \hline
    \input{md4.benchmarks.tex} \\
    \hline
  \end{tabular}
  \caption{Relative speedup (see text for details) of the \code{LibintX} GPU MD code vs the reference \code{Libint} OSHGP CPU code.}
  \label{tab:speedup4}
\end{table}

\begin{table}
  \centering
  \begin{tabular}{c|ccc}
    \hline
    Integral Class & \multicolumn{3}{c}{$\{K_\mathrm{bra},K_\mathrm{ket}\}$}  \\
    & $\{1,1\}$ & $\{1,5\}$ & $\{5,5\}$ \\
    \hline
    \input{md3.benchmarks.tex} \\
    \hline
  \end{tabular}
  \caption{Relative speedup (see text for details) of the \code{LibintX} GPU MD code vs the reference \code{Libint} OSHGP CPU code for 3-center integrals.}
  \label{tab:speedup3}
\end{table}

\begin{table}
  \centering
  \begin{tabular}{c|ccc}
    \hline
    Integral Class & \multicolumn{3}{c}{$\{K_\mathrm{bra},K_\mathrm{ket}\}$}  \\
    & $\{1,1\}$ & $\{1,5\}$ & $\{5,5\}$ \\ \hline
    \input{md4.rate.tex} \\
    \hline
  \end{tabular}
  \caption{Absolute rate of evaluation of 4-center integrals (shellsets/second) on single NVIDIA V100 GPU.}
  \label{tab:nps4}
\end{table}

\begin{table}
  \centering
  \begin{tabular}{c|ccc}
    \hline
    Integral Class & \multicolumn{3}{c}{$\{K_\mathrm{bra},K_\mathrm{ket}\}$}  \\
    & $\{1,1\}$ & $\{1,5\}$ & $\{5,5\}$ \\ \hline
    \input{md3.rate.tex} \\
    \hline
  \end{tabular}
  \caption{Absolute rate of evaluation of 3-center integrals (shellsets/second) on single NVIDIA V100 GPU.}
  \label{tab:nps3}
\end{table}

\Cref{tab:nps4,tab:nps3} reports the effective throughput of the \code{LibintX} MD engine (shellsets per second) for 4- and 3-center integrals, respectively. This data can be used by other researchers for performance assessment relative to \code{LibintX} without having to run it themselves.

\section{Preliminary Assessment: Exchange Operator Construction}
\label{sec:exchange}

To assess the usability of the AO integral engine for more complex tasks, we developed a prototype engine for evaluation of the exchange operator:
\begin{align}
K_{bc} = \sum_{ad} [ab|cd] D_{ad},
\end{align}
where $D_{ad}$ are the matrix elements of the (1-particle) density matrix.

Contraction of density $D$ to produce exchange operator $K$ is memory bound by reading $[ab|cd]$ from the main GPU memory. Given integrals $V[ij,ab,cd,kl]$ and density matrix $D[ad,il]$ (and its permutations), rather than transposing integral, much smaller density matrices are permuted and broadcast to match the integral layout:
\begin{eqnarray}
\label{eq:K-eq1}
D[ij,ad,l] & = & D[ad,il], \\
\label{eq:K-eq2}
K[ij,bc,k] & = & V[ij,ab,cd,kl] D[ij,ad,l], \\
\label{eq:K-eq3}
K[j,bc,k] & = & \sum_j K[ij,bc,k].
\end{eqnarray}
In step \eqref{eq:K-eq1} the density batch is broadcast to all $j$ for which there are non-negligible $ij$ shell pairs.
Step \eqref{eq:K-eq2} is a batched tensor contraction ($ij$ is the batching index) where the vast majority of arithmetic work is performed. The last step is a reduction over $i$. Making sure that step \eqref{eq:K-eq2} has perfect data parallelism over $ij$ guarantees maximum bandwidth {\it and} defers atomic reductions over $i$ until the much cheaper last step.
Reduction over $l$ can be likewise deferred but in our experience atomic reductions over $l$ have negligible overhead.

Our exchange implementation is not yet complete or fully optimized.
We report these preliminary timings for illustration only, to show one use case of \code{LibintX} and its projected performance.
We postponed making the source code public until the work is finished and API is finalized.

\Cref{tab:6-31g*} reports exchange evaluating timings for a standard set of  molecules used to benchmark performance of the SCF solvers and its components in the literature.\cite{VRG:ufimtsev:2009:JCTC,VRG:laqua:2020:JCTC,VRG:barca:2021:JCTC,VRG:williams-young:2023:JCP} Due to our choice regarding solid harmonic AOs, the Cartesian Pople basis sets basis are treated as pure spherical basis rather than Cartesian; we denote such basis variant by adding suffix ``(5D)''.

\begin{table}
  \centering
  \begin{tabular}{c|rddr}
    \hline
    Molecule$^a$ & $N_\mathrm{AO}$ & \multicolumn{1}{c}{$t_\mathrm{i}$} & \multicolumn{1}{c}{$t_\mathrm{d}$} & $N_{ijkl}$ \\
    \hline
    \input{exchange-6-31g_d_.tex} \\ \hline
  \end{tabular}
  \\
    $^a$ Geometries were obtained from Refs. \onlinecite{VRG:maurer:2012:JCP,VRG:williams-young:2020:FC,VRG:manathunga:2021:JCTC,VRG:barca:2021:JCTC}.

  \caption{The exchange operator timings for a standard benchmark molecules using the 6-31g*(5D) AO basis. $t_\mathrm{i}$ and $t_\mathrm{d}$ report time (s) spent in the integral evaluation and digestion, respectively. $N_\mathrm{AO}$ and $N_\mathrm{ijkl}$ are the number of AO functions and the total number of shellsets evaluated.}
  \label{tab:6-31g*}

\end{table}

\Cref{tab:taxol} reports timings for exchange evaluation of a single taxol molecule in def2 and augmented correlation-consistent basis sets ranging from double- to quadruple-zeta. As expected, increasing the cardinal number by 1 raises the cost of exchange evaluation by roughly an order of magnitude; but the fact that high-$l$ basis sets can be used for practical computation on the GPU without additional approximations make these computations possible.  The higher contraction degree of correlation-consistent basis sets compared to their def2 counterparts leads to significantly higher cost per quartet for the former.

\begin{table}
  \centering
  \begin{tabular}{c|rddr}
    \hline
    Basis & $N_\mathrm{AO}$ & \multicolumn{1}{c}{$t_{\rm i}$} & \multicolumn{1}{c}{$t_{\rm d}$} & $N_{ijkl}$ \\
    \hline
    \input{exchange-taxol.tex} \\ \hline
  \end{tabular}
  \caption{The exchange operator timings for taxol molecule using double-, triple-, and quadruple-zeta basis sets of two popular basis set families. $t_\mathrm{i}$ and $t_\mathrm{d}$ report time (s) spent in the integral evaluation and digestion, respectively. $N_\mathrm{AO}$ and $N_\mathrm{ijkl}$ are the number of AO functions and the total number of shellsets evaluated.}
  \label{tab:taxol}
\end{table}

Large computations with over 20,000 basis functions are possible with this preliminary naive implementation on a single 16GB V100 GPU. For example, a single exchange matrix evaluation for ubiquitin in def2-TZVP basis, comprised of 22,442 AOs, took $\sim6000$ seconds involving $1.674\times10^{12}$ non-negligible integral shellsets.

\section{Summary}
\label{sec:summary}

Despite being more expensive than the alternatives, the simplicity and regularity of the matrix form of the McMurchie-Davidson (MD) scheme makes it a better candidate than complex recurrence-based schemes for evaluation of 2-particle integrals over Gaussian AOs on modern GPUs. This paper extends our original work\cite{VRG:asadchev:2023:JPCA} with new refinements aimed at lower angular momenta as well as specializations for the three-center integrals. Preliminary evaluation of the exchange matrix is also presented and assessed for a range of systems with up to more than 22,000 basis functions and up to quadruple-zeta basis sets. The computer implementation of these algorithms is distributed in the open-source C++ {\tt LibintX} library under
LGPL3; it is available publicly at \textbf{\url{github.com:ValeevGroup/LibintX}}.
We welcome contributions to \code{LibintX} from other researchers.

Future developments will focus on developing similar methods for standard CPUs with wide SIMD units, finalizing the exchange operator code, introducing support for geometrical derivatives, and post-SCF methods.

\begin{acknowledgments}
This work was supported by the U.S. Department of Energy award DE-SC0022263 provided via the Scientific Discovery through Advanced Computing (SciDAC) program by the Offices of Advanced Scientific Computing Research (ASCR) and Basic Energy Sciences (BES). We also acknowledge Advanced Research Computing at Virginia Tech (www.arc.vt.edu) for providing computational resources and technical support that have contributed to the results reported within this paper.
\end{acknowledgments}

\bibliography{vrgrefs}

\end{document}

%% file: md4.benchmarks.tex
$[11|00]$ & 592 & 796 & 604 \\
$[22|00]$ & 197 & 335 & 221 \\
$[33|00]$ & 107 & 197 & 124 \\
$[44|00]$ & 121 & 196 & 146 \\
$[55|00]$ & 147 & 162 & 128 \\
$[66|00]$ & 176 & 212 & 151 \\
$[10|10]$ & 643 & 756 & 585 \\
$[20|20]$ & 470 & 511 & 399 \\
$[30|30]$ & 113 & 183 & 150 \\
$[40|40]$ & 250 & 385 & 362 \\
$[50|50]$ & 133 & 214 & 219 \\
$[60|60]$ & 115 & 184 & 195 \\
$[00|00]$ & 593 & 1084 & 1015 \\
$[11|11]$ & 156 & 252 & 206 \\
$[22|22]$ & 223 & 222 & 165 \\
$[33|33]$ & 268 & 165 & 117 \\
$[44|44]$ & 300 & 167 & 119 \\
$[55|55]$ & 603 & 279 & 176 \\
$[66|66]$ & 1171 & 434 & 194 \hspace{-7pt}

%% file: md3.benchmarks.tex
$[1|00]$ & 1162 & 1629 & 1492  \\
$[2|00]$ & 816 & 1112 & 1043  \\
$[3|00]$ & 698 & 819 & 869  \\
$[4|00]$ & 571 & 775 & 747  \\
$[5|00]$ & 531 & 633 & 586  \\
$[6|00]$ & 445 & 467 & 401  \\
$[0|11]$ & 537 & 935 & 846  \\
$[0|22]$ & 274 & 338 & 261  \\
$[0|33]$ & 132 & 172 & 124  \\
$[0|44]$ & 148 & 72 & 53  \\
$[0|55]$ & 121 & 54 & 43  \\
$[0|66]$ & 182 & 86 & 142  \\
$[0|00]$ & 1026 & 1768 & 1719  \\
$[1|11]$ & 225 & 576 & 474  \\
$[2|22]$ & 212 & 133 & 97  \\
$[3|33]$ & 180 & 81 & 100  \\
$[4|44]$ & 237 & 126 & 227  \\
$[5|55]$ & 385 & 206 & 369  \\
$[6|66]$ & 325 & 175 & 265 \hspace{-7pt}

%% file: md4.rate.tex
$[11|00]$ & \num{1.98e+09} & \num{1.28e+09} & \num{3.48e+08} \\
$[22|00]$ & \num{4.75e+08} & \num{3.98e+08} & \num{9.11e+07} \\
$[33|00]$ & \num{1.40e+08} & \num{1.33e+08} & \num{2.72e+07} \\
$[44|00]$ & \num{5.42e+07} & \num{4.11e+07} & \num{8.68e+06} \\
$[55|00]$ & \num{3.08e+07} & \num{1.66e+07} & \num{3.76e+06} \\
$[66|00]$ & \num{1.82e+07} & \num{1.12e+07} & \num{2.32e+06} \\
$[10|10]$ & \num{2.10e+09} & \num{1.23e+09} & \num{3.31e+08} \\
$[20|20]$ & \num{1.01e+09} & \num{5.17e+08} & \num{1.27e+08} \\
$[30|30]$ & \num{1.28e+08} & \num{8.19e+07} & \num{1.75e+07} \\
$[40|40]$ & \num{8.31e+07} & \num{3.62e+07} & \num{7.52e+06} \\
$[50|50]$ & \num{1.80e+07} & \num{7.19e+06} & \num{1.56e+06} \\
$[60|60]$ & \num{6.71e+06} & \num{2.47e+06} & \num{5.43e+05} \\
$[00|00]$ & \num{4.53e+09} & \num{2.98e+09} & \num{9.11e+08} \\
$[11|11]$ & \num{2.59e+08} & \num{2.05e+08} & \num{5.27e+07} \\
$[22|22]$ & \num{2.25e+07} & \num{1.04e+07} & \num{2.12e+06} \\
$[33|33]$ & \num{2.47e+06} & \num{8.00e+05} & \num{1.69e+05} \\
$[44|44]$ & \num{3.25e+05} & \num{9.59e+04} & \num{2.05e+04} \\
$[55|55]$ & \num{1.11e+05} & \num{3.14e+04} & \num{6.73e+03} \\
$[66|66]$ & \num{3.47e+04} & \num{9.74e+03} & \num{1.99e+03} \hspace{-7pt}

%% file: md3.rate.tex
$[1|00]$ & \num{5.18e+09} & \num{3.29e+09} & \num{1.09e+09}  \\
$[2|00]$ & \num{3.30e+09} & \num{2.10e+09} & \num{7.20e+08}  \\
$[3|00]$ & \num{2.52e+09} & \num{1.44e+09} & \num{5.61e+08}  \\
$[4|00]$ & \num{1.72e+09} & \num{1.13e+09} & \num{3.70e+08}  \\
$[5|00]$ & \num{1.34e+09} & \num{8.15e+08} & \num{2.64e+08}  \\
$[6|00]$ & \num{9.37e+08} & \num{5.17e+08} & \num{1.46e+08}  \\
$[0|11]$ & \num{2.17e+09} & \num{1.72e+09} & \num{5.47e+08}  \\
$[0|22]$ & \num{7.81e+08} & \num{4.48e+08} & \num{1.17e+08}  \\
$[0|33]$ & \num{2.04e+08} & \num{1.27e+08} & \num{2.90e+07}  \\
$[0|44]$ & \num{7.05e+07} & \num{1.57e+07} & \num{3.23e+06}  \\
$[0|55]$ & \num{2.69e+07} & \num{5.77e+06} & \num{1.29e+06}  \\
$[0|66]$ & \num{1.97e+07} & \num{4.67e+06} & \num{2.20e+06}  \\
$[0|00]$ & \num{9.63e+09} & \num{5.15e+09} & \num{1.64e+09}  \\
$[1|11]$ & \num{7.25e+08} & \num{8.88e+08} & \num{2.59e+08}  \\
$[2|22]$ & \num{1.75e+08} & \num{5.43e+07} & \num{1.18e+07}  \\
$[3|33]$ & \num{3.34e+07} & \num{7.48e+06} & \num{2.67e+06}  \\
$[4|44]$ & \num{8.88e+06} & \num{1.90e+06} & \num{8.57e+05}  \\
$[5|55]$ & \num{4.17e+06} & \num{8.91e+05} & \num{3.98e+05}  \\
$[6|66]$ & \num{1.18e+06} & \num{2.63e+05} & \num{1.02e+05}
\hspace{-7pt}

%% file: exchange-6-31g_d_.tex
taxol     & 970  &  3.6  &  0.4 &    983,376,040 \\
olestra   & 3,006 &  8.5  &  1.6 &   3,838,322,047 \\
(Gly)$_{120}$    & 7,458 & 30.1  &  6.2 &  11,387,587,514 \\
(AT)$_{8}$      & 4,968 & 59.2  & 12.3 &  23,002,026,695 \\
crambin   & 5,232 & 77.6  & 18.1 &  36,253,672,158 \\
ubiquitin & 9,690 & 234.8 & 63.3 & 127,407,939,457 \hspace{-7pt}

%% file: exchange-taxol.tex
def2-SVP    & 1,099 & 3.1    & 0.5   &  1,142,492,179  \\
def2-TZVP   & 2,185 & 32.4   & 5.1   &  7,077,935,024 \\
def2-QZVP   & 4,947 & 563.6  & 70.6  & 45,762,486,375 \\ \hline
aug-cc-pVDZ & 1,490 & 40.3   & 4.6   &  6,030,285,469 \\
aug-cc-pVTZ & 2,834 & 405.4  & 34.2  & 18,791,062,573 \\
aug-cc-pVQZ & 4,618 & 3539.7 & 142.6 & 42,381,637,076 \hspace{-7pt}